\documentclass{article}
\usepackage{amsmath}
\usepackage{amsfonts}
\usepackage{amssymb}

\input{tcilatex}

\begin{document}

\title{Spinorial Field and Lyra Geometry}
\author{R. Casana, C. A. M. de Melo and B. M. Pimentel \\
{\small Instituto de F\'{\i}sica Te\'orica, Universidade Estadual Paulista}\\
[-0.1cm] {\small Rua Pamplona 145, CEP 01405-900, S\~ao Paulo, SP, Brazil}}
\maketitle

\begin{abstract}
The Dirac field is studied in a Lyra space-time background by means of the
classical Schwinger Variational Principle. We obtain the equations of
motion, establish the conservation laws, and get a scale relation relating
the energy-momentum and spin tensors. Such scale relation is an intrinsic
property for matter fields in Lyra background.
\end{abstract}


\section{ Introduction}

After Einstein's approach to gravitation \cite{Einstein}, several others
theories have been developed, as part of efforts to cure problems arising
when the gravitational field is coupled to matter fields. Thus, as soon as
Einstein presented the General Relativity, Weyl \cite{Weyl} proposed a new
geometry in which a new scalar field accompanies the metric field and
changes the scale of length measurements. The aim was to unify gravitation
and electromagnetism, but this theory was briefly refuted by Einstein
because the non-metricity had direct consequences over the spectral lines of
elements which has never been observed.

The study of the spin coupling to gravitation has been and it is a central
problem. The principal path to incorporate spin in geometrical theories of
gravitation is the use of so called Riemann-Cartan geometry \cite{R-C} which
has a nonsymmetric connection, in such a way that a new geometrical concept
enters in scene: the torsion. However, analyzing the Cauchy data, one can
proof that the torsion is a non-propagating entity and therefore must be
different of zero only in the interior of matter.

Some years after Weyl, Lyra \cite{Lyra} proposed a new geometry which
removes the non-integrability of the length transfer of a vector under
parallel transport (the metricity condition is restored) by introducing a
gauge or scale function into the modified Riemannian geometry. The study of
this geometry was completed by Scheibe \cite{Lyra} and, it was analyzed by
Sen \cite{Sen} and several others as an alternative to describe the
gravitational field, and more recently it has been applied to study viscous
and higher dimensional \cite{ViscLyra} cosmological models, domain walls 
\cite{DomWall}, and several others applications. The attractive of Lyra's
geometry resides in the fact that the torsion is propagating which in the
context of spin-gravitational coupling is interesting.

Following the spirit of finding a solution to the difficulties \cite{ref1}
encountered in the quantization of Einstein's theory of gravitation, many
endeavours have been made to generalize and modify such theory. Among them
some may prove to be of crucial importance in the formulation of a future
and successful quantum gravity as for example the torsion \cite{ref2},
conformal gauge symmetry \cite{Weyl,ref3} and supersymmetry \cite{ref4}.

On the other hand, in the early days of quantum field theory, many
calculations were undertaken in which the electromagnetic field was
considered as a classical background field interacting with quantized
matter. Such semiclassical approach readily gives results in complete
accordance with the fully quantized theory. And, since the quantization of
gravity is still a controversial matter, semiclassical behavior of quantum
matter in non-Euclidean manifolds appears as natural way to achieve some
results about the fully quantized theory again. In this context, the role of
symmetries on the manifold is of special importance in the definition of the
particle content \cite{Birrel-Davies}.

In this paper we want to use the Lyra Geometry, a curved and torsioned
manifold where scale symmetry is implemented in a non-usual way, to study
the behavior of the spinorial (Dirac) field in presence of a classical
background. The main tool for this study will be the Schwinger Variational
Principle, which is able to construct, in a unique and direct proceeding,
equations of motion, canonical generators and conservation laws. Such
properties were studied in the context of Riemann-Cartan geometry in \cite%
{Nieh-1}

In the next section the basics of Lyra Geometry \cite{Lyra,Sen} are
introduced, followed by the definition of covariant differentiation of
spinors in this manifold. Next, the classical Schwinger Action Principle 
\cite{QuanField, Sudarshan} is presented and then applied in section \ref%
{Contas}. Finally, we make some comments on the results obtained.

\section{Lyra Geometry}

Lyra geometry is a generalization of the usual Riemannian one, constructed
by attaching to each coordinate system a \emph{scale function} $\phi \left(
x\right) $. Let be $M\subseteq \mathbb{R}^{N}$ and $U$ an open set of $%
\mathbb{R}^{n}$, also let be $\chi :U\curvearrowright M$. Since $\chi $\ is
injective ($\curvearrowright $), is settled a univalent equivalence relation
between points in the domain $U$ and the image $\chi \left( U\right) $. The
couple $\left( U,\chi \right) $ is called a coordinate system. A \emph{%
reference system} is defined by the triple $\left( U,\chi ,\phi \right) $
such that 
\begin{equation}
\bar{\phi}\left( \bar{x}\right) =\bar{\phi}\left( x\left( \bar{x}\right)
;\phi \left( x\left( \bar{x}\right) \right) \right) \qquad ,\qquad \frac{%
\partial \bar{\phi}}{\partial \phi }\not=0\,.
\end{equation}%
Therefore, the scale function does not transform like a scalar function, but
like a diffeomorphism in the same way as a coordinate. Then, by general
changes of the reference system a Lyra vector transforms as 
\begin{equation}
\bar{A}^{\mu }\left( \bar{x}\right) =\frac{\bar{\phi}\left( \bar{x}\right) }{%
\phi \left( x\right) }\frac{\partial \bar{x}^{\mu }}{\partial x^{\nu }}%
A^{\nu }\left( x\right) \,.
\end{equation}

In differential geometry, parallel transport defined as a \emph{linear} map
from vectors in a given point to vectors in another infinitesimal neighbor
point, on the same manifold, 
\begin{equation}
\delta V^{\mu }\equiv L^{\mu }{}_{\alpha }\left( V\right) dx^{\alpha }\,.
\end{equation}%
Besides the linearity, 
\begin{equation}
\delta \left( \lambda V_{1}^{\mu }+V_{2}^{\mu }\right) =\lambda L^{\mu
}{}_{\alpha }\left( V_{1}\right) dx^{\alpha }+L^{\mu }{}_{\alpha }\left(
V_{2}\right) dx^{\alpha }\,,
\end{equation}%
one also demands a Leibnitz composition rule, 
\begin{equation}
\delta \left( V_{1}^{\mu }V_{2}^{\nu }\right) =L^{\mu }{}_{\alpha }\left(
V_{1}\right) V_{2}^{\nu }dx^{\alpha }+L^{\nu }{}_{\alpha }\left(
V_{2}\right) V_{1}^{\mu }dx^{\alpha }\,.
\end{equation}%
Notwithstanding, the crucial point in the definition of the parallel
transport is on the linearity, dictated by 
\begin{equation}
L^{\mu }{}_{\alpha }\left( \lambda V\right) =\lambda L^{\mu }{}_{\alpha
}\left( V\right)
\end{equation}%
which, in conjunction with the infinitesimal nature of the transformation,
conducts to conclusion the map is proportional to the transported vector
itself\footnote{%
It is important to note this is a general expression, and $\Gamma $ is the
connection on the manifold under study, and not just the Lyra connection
defined bellow.}, i.e., 
\begin{equation}
\delta V^{\mu }=-\Gamma ^{\mu }{}_{\alpha \beta }V^{\beta }dx^{\alpha }\,.
\end{equation}%
Actually, it is simple to proof that \emph{all} linear infinitesimal
application has this property.

With these considerations, the Lyra parallel transportation is defined by 
\begin{equation}
\delta A^{\mu}\left( x\right) \equiv-\Gamma^{\mu}{}_{\alpha\beta}A^{\beta
}\phi\left( x\right) dx^{\alpha}\,,  \label{DefConexLyra}
\end{equation}
where, 
\begin{equation}
\Gamma^{\rho}{}_{\mu\nu}=\frac{1}{\phi}\mathring{\Gamma}^{\rho}{}_{\mu\nu }+%
\frac{1}{\phi}\left[ \partial_{\nu}\ln\left( \frac{\phi}{\bar{\phi}}\right)
\delta_{\,\mu}^{\rho}-\partial_{\sigma}\ln\left( \frac{\phi}{\bar{\phi}}%
\right) g_{\nu\mu}g^{\rho\sigma}\right] \,,  \label{ConexLyra}
\end{equation}
with 
\begin{equation}
\mathring{\Gamma}^{\rho}{}_{\mu\nu}\equiv\frac{1}{2}g^{\rho\sigma}\left(
\partial_{\mu}g_{\nu\sigma}+\partial_{\nu}g_{\sigma\mu}-\partial_{\sigma
}g_{\mu\nu}\right) \,.  \label{Christoffel}
\end{equation}

Under a general reference system transformation, $\Gamma$ transforms as 
\begin{equation}
\Gamma^{\rho}{}_{\alpha\beta}=\frac{\bar{\phi}}{\phi}\bar{\Gamma}%
^{\nu}{}_{\lambda\varepsilon}\frac{\partial x^{\rho}}{\partial\bar{x}^{\nu}}%
\frac{\partial\bar{x}^{\lambda}}{\partial x^{\alpha}}\frac{\partial\bar {x}%
^{\varepsilon}}{\partial x^{\beta}}+\frac{1}{\phi}\frac{\partial x^{\rho}}{%
\partial\bar{x}^{\nu}}\frac{\partial^{2}\bar{x}^{\nu}}{\partial x^{\alpha
}\partial x^{\beta}}+\frac{1}{\phi}\frac{\partial}{\partial x^{\alpha}}%
\ln\left( \frac{\bar{\phi}}{\phi}\right) \delta_{\,\beta}^{\rho}\,.
\label{TransConexLyra}
\end{equation}

Using the standard definitions, one finds for the Lyra geometry the \emph{%
covariant derivative} 
\begin{equation}
\nabla _{\mu }F_{\,\,\sigma \ldots }^{\lambda \ldots }\equiv \frac{1}{\phi }%
\partial _{\mu }F_{\,\,\sigma \ldots }^{\lambda \ldots }+\Gamma ^{\lambda
}{}_{\mu \alpha }F_{\,\,\sigma \ldots }^{\alpha \ldots }+\,\ldots
\,-\,\Gamma ^{\alpha }{}_{\mu \sigma }F_{\,\alpha \ldots \,}^{\lambda \ldots
}-\,\ldots \,,
\end{equation}%
for a general Lyra tensor. We also find the Lyra \emph{curvature tensor} 
\begin{equation}
R^{\rho }{}_{\beta \alpha \sigma }\equiv \frac{1}{\phi ^{2}}\left( \frac{%
\partial \left( \phi \Gamma ^{\rho }{}_{\alpha \sigma }\right) }{\partial
x^{\beta }}-\frac{\partial \left( \phi \Gamma ^{\rho }{}_{\beta \sigma
}\right) }{\partial {x}^{\alpha }}+\phi \Gamma ^{\rho }{}_{\beta \lambda
}\phi \Gamma ^{\lambda }{}_{\alpha \sigma }-\phi \Gamma ^{\rho }{}_{\alpha
\lambda }\phi \Gamma ^{\lambda }{}_{\beta \sigma }\right) ,  \label{CurvLyra}
\end{equation}%
and the Lyra \emph{torsion tensor} 
\begin{equation}
\tau _{\alpha \beta }{}^{\mu }\equiv 2\left( \Gamma ^{\mu }{}_{\left[ \alpha
\beta \right] }+\frac{1}{\phi }\delta _{\,[\beta }^{\mu }\partial _{\alpha
]}\ln \phi \right) .  \label{TorcLyra}
\end{equation}

The antisymmetrization is defined by $\Gamma^{\mu}{}_{\left[ \alpha \beta%
\right] }\equiv\frac{1}{2}\left(
\Gamma^{\mu}{}_{\alpha\beta}-\Gamma^{\mu}{}_{\beta\alpha}\right) $, and the
metricity condition $\nabla_{\lambda}g_{\mu\nu}=0$ imply 
\begin{equation}
\Gamma^{\rho}{}_{\mu\nu}=\frac{1}{\phi}\mathring{\Gamma}^{\rho}{}_{\mu\nu
}+\Gamma^{\rho}{}_{\left[ \mu\nu\right] }-\Gamma^{\alpha}{}_{\left[ \mu\sigma%
\right] }g_{\nu\alpha}g^{\rho\sigma}-\Gamma^{\alpha}{}_{\left[ \nu\sigma%
\right] }g_{\alpha\mu}g^{\rho\sigma}\,.  \label{ConexST}
\end{equation}

In the next section we introduce the behavior of spinorial fields in the
Lyra geometry.

\section{Covariant Differentiation of Spinors\label{CovDif}}

To construct the covariant derivative of Dirac field in Lyra geometry, we
follow the standard procedure of analyzing the behavior of the field under
local Lorentz transformations, 
\begin{equation}
\psi\left( x\right) \rightarrow\psi^{\prime}\left( x\right) =U\left( L\left(
x\right) \right) \psi\left( x\right) \,,  \label{LocalLorentz}
\end{equation}
where $U$\ is an one-half spin representation of Lorentz group.

What we want is to define a \emph{spin connection} $S_{\mu }$ in a such way
that the object 
\begin{equation}
\nabla _{\mu }\psi \equiv \frac{1}{\phi }\partial _{\mu }\psi +S_{\mu }{}\psi
\label{cov-fer}
\end{equation}%
transforms like a spinor, 
\begin{equation}
\nabla _{\mu }\psi \rightarrow \left( \nabla _{\mu }\psi \right) ^{\prime
}=U\left( x\right) \nabla _{\mu }\psi
\end{equation}%
and therefore $S$ transforms as 
\begin{equation}
S_{\mu }^{\prime }=U\left( x\right) S_{\mu }U^{-1}\left( x\right) -\frac{1}{%
\phi }\left( \partial _{\mu }U\right) U^{-1}\left( x\right) \,.
\label{TransfSpin}
\end{equation}

If we remember the Dirac matrices algebra 
\begin{equation}
\left\{ \gamma^{\mu},\gamma^{\nu}\right\} =2g^{\mu\nu}
\end{equation}
and use the requirement of compatibility among spin connection and metric,
we can found 
\begin{equation}
\nabla_{\mu}g^{\alpha\beta}=0\Rightarrow\nabla_{\mu}\left\{ \gamma^{\alpha
},\gamma^{\beta}\right\} =0\,.  \label{metrica-spin}
\end{equation}

From the covariant derivative of the fermion field (\ref{cov-fer}) and
remembering that $\bar{\psi}\psi$ must be a scalar under the transformation (%
\ref{LocalLorentz}), it follows that 
\begin{equation}
\nabla_{\mu}\bar{\psi}=\frac{1}{\phi}\partial_{\mu}\bar{\psi}-\bar{\psi}%
S_{\mu}\,.
\end{equation}

Then, we use the covariant derivative of the fermionic current 
\begin{equation*}
\nabla_{\mu}\left( \bar{\psi}\gamma^{\nu}\psi\right) =\frac{1}{\phi}%
\partial_{\mu}\left( \bar{\psi}\gamma^{\nu}\psi\right)
+\Gamma^{\nu}{}_{\mu\lambda}\left( \bar{\psi}\gamma^{\lambda}\psi\right)
=\left( \nabla_{\mu}\bar{\psi}\right) \gamma^{\nu}\psi+\bar{\psi}\left(
\nabla_{\mu }\gamma^{\nu}\right) \psi+\bar{\psi}\gamma^{\nu}\left(
\nabla_{\mu}\psi\right)
\end{equation*}
to get the following expression for the covariant derivative\footnote{%
This defines the general covariant derivative of an object with both
tensorial and spinorial indexes, 
\begin{equation*}
\nabla_{\mu}F^{\lambda}{}_{B}{}^{A}=\frac{1}{\phi}\partial_{\mu}F^{%
\lambda}{}_{B}{}^{A}+\Gamma^{\lambda}{}_{\mu\alpha}F^{%
\alpha}{}_{B}{}^{A}+S_{\mu}{}^{A}{}_{C}F^{\lambda}{}_{B}{}^{C}-S_{%
\mu}{}^{C}{}_{B}F^{\lambda}{}_{C}{}^{A}.
\end{equation*}%
} of $\gamma^{\nu }$ 
\begin{equation}
\nabla_{\mu}\gamma^{\nu}=\frac{1}{\phi}\partial_{\mu}\gamma^{\nu}+\Gamma^{%
\nu
}{}_{\mu\lambda}\gamma^{\lambda}+S_{\mu}\gamma^{\nu}-\gamma^{\nu}S_{\mu}\,.
\end{equation}
A sufficient condition compatible with (\ref{metrica-spin}) is 
\begin{equation}
\nabla_{\mu}\gamma^{\nu}=\frac{1}{\phi}\partial_{\mu}\gamma^{\nu}+\Gamma^{%
\nu
}{}_{\mu\lambda}\gamma^{\lambda}+S_{\mu}\gamma^{\nu}-\gamma^{\nu}S_{\mu}=0\,.
\label{cov-gamma}
\end{equation}

To solve the equation above we introduce the tetrad field $e^{\mu }{}_{a}$
and its inverse $e_{\mu }{}^{a}$ related to the space-time metric by the
following equations 
\begin{eqnarray}
g^{\mu \nu }(x) &=&\eta ^{ab}\,e^{\mu }{}_{a}(x)e^{\nu }{}_{b}(x)\,,  \notag
\\
g_{\mu \nu }(x) &=&\eta _{ab}e_{\mu }{}^{a}(x)e_{\nu }{}^{b}(x)\,, \\
e_{\mu }{}^{a}(x) &=&g_{\mu \nu }(x)\eta ^{ab}e^{\nu }{}_{b}(x)\,,  \notag
\end{eqnarray}%
where $\det \left( e_{\mu }{}^{a}\right) =e=\sqrt{-g}$.

Therefore, the metricity condition can be expressed as 
\begin{equation}
\nabla _{\mu }e^{\nu }{}_{a}=\frac{1}{\phi }\partial _{\mu }e^{\nu
}{}_{a}+\Gamma ^{\nu }{}_{\mu \lambda }e^{\lambda }{}_{a}+{}\omega _{\mu
a}{}^{b}e^{\nu }{}_{b}=0\,.  \label{tetrad-metric}
\end{equation}%
Expressing $\gamma ^{\nu }$ in terms of the tetrad fields, $\gamma ^{\nu
}=e^{\nu }{}_{a}\gamma ^{a}$, in the equation (\ref{cov-gamma}) we get 
\begin{equation}
\omega _{\mu a}{}^{b}e^{\nu }{}_{b}\gamma ^{a}=S_{\mu }\gamma ^{\nu }-\gamma
^{\nu }S_{\mu }=\left[ S_{\mu },\gamma ^{\nu }\right] \,,
\end{equation}%
from which we found that 
\begin{equation}
S_{\mu }=\frac{1}{2}\omega _{\mu ab}\Sigma ^{ab}\,,
\end{equation}%
where we define 
\begin{equation}
\Sigma ^{ab}=\frac{1}{4}\left[ \gamma ^{a},\gamma ^{b}\right] \,.
\end{equation}

\section{Schwinger Action Principle}

In following we use a classical version of the Schwinger Action Principle
such as it was used in the context of Classical Mechanics by Sudarshan and
Mukunda \cite{Sudarshan}. The Schwinger Action Principle is the most general
version of the usual variational principles. It was proposed originally at
the scope of the Quantum Field Theory \cite{QuanField}, but its application
goes beyond this area. In the classical context, the basic statement of the
Schwinger Principle is%
\begin{equation*}
\delta S=\delta \int_{\Omega }dxe\phi ^{4}\mathcal{L}=\int_{\partial \Omega
}d\sigma _{\mu }G^{\mu }~,
\end{equation*}%
where $S$ is the classical action and $G^{\mu }$ are the generators of the
canonical transformations. The Schwinger Principle can be employed, choosing
suitable variations in each case, to obtain commutation relations in the
quantum context or canonical transformations in the classical one, as well
as equations of motion or still perturbative expansions.

Here, we will apply the Action Principle to derive equations of motion of
the Dirac field in an external Lyra background and its conservations laws
associated with translations and rotations in such space.

\section{Dynamics of the Dirac Field Coupled to the Lyra Manifold\label%
{Contas}}

With the definitions of section \ref{CovDif} the Dirac Lagrangian in
Minkowski spacetime, expressed by 
\begin{equation}
\mathcal{L}^{M_{4}}=\frac{i}{2}\left( \bar{\psi}\gamma ^{a}\partial _{a}\psi
-\partial _{a}\bar{\psi}\gamma ^{a}\psi \right) -m\bar{\psi}\psi ~,
\label{MinkAction}
\end{equation}%
can be generalized by the minimal coupling procedure to the Lyra space-time
as 
\begin{equation}
S=\int_{\Omega }dx\phi ^{4}e\left\{ \frac{i}{2}\bar{\psi}\gamma ^{\mu
}\nabla _{\mu }\psi -\frac{i}{2}\nabla _{\mu }\bar{\psi}\gamma ^{\mu }\psi -m%
\bar{\psi}\psi \right\} \,.  \label{lag-cv}
\end{equation}

To derive in a direct way the equations of motion, energy-momentum and spin
tensor for the Dirac field coupled to Lyra manifold we will use the
Schwinger action principle. Thus, making the variation of the action
integral we get 
\begin{eqnarray}
\delta S &=&\!\!\int_{\Omega }dx~\phi ^{4}e\left\{ \frac{4}{\phi }\mathcal{L}%
-\frac{i}{2\phi ^{2}}\bar{\psi}\gamma ^{\mu }\partial _{\mu }\psi +\frac{i}{%
2\phi ^{2}}\partial _{\mu }\bar{\psi}\gamma ^{\mu }\psi \right\} \delta \phi
~~+~\int_{\Omega }dx~\phi ^{4}e\left( \frac{\delta e}{e}\right) \mathcal{L}%
~\ +  \notag \\
&&+\int_{\Omega }dx~\phi ^{4}e\frac{i}{2}\left\{ \frac{{}}{{}}\bar{\psi}%
\left( \delta \gamma ^{\mu }\right) \nabla _{\mu }\psi -\nabla _{\mu }\bar{%
\psi}\left( \delta \gamma ^{\mu }\right) \psi \right\} ~~+~\int_{\Omega
}dx~\phi ^{4}e\frac{i}{2}\left\{ \frac{{}}{{}}\bar{\psi}\gamma ^{\mu }\left(
\delta S_{\mu }\right) \psi +\bar{\psi}\left( \delta S_{\mu }\right) \gamma
^{\mu }\psi \right\} +  \notag \\
&&+\int_{\Omega }dx~\phi ^{4}e\left\{ \frac{i}{2}\left[ \left( \delta \bar{%
\psi}\right) \gamma ^{\mu }\nabla _{\mu }\psi -\left( \frac{1}{\phi }\delta
\left( \partial _{\mu }\bar{\psi}\right) -\left( \delta \bar{\psi}\right)
S_{\mu }\right) \gamma ^{\mu }\psi \right] -m\left( \delta \bar{\psi}\right)
\psi \right\} +  \label{action-variation} \\
&&\!+\int_{\Omega }dx~\phi ^{4}e\left\{ \frac{i}{2}\left[ \bar{\psi}\gamma
^{\mu }\left( \frac{1}{\phi }\delta \left( \partial _{\mu }\psi \right)
+S_{\mu }\left( \delta \psi \right) \right) -\nabla _{\mu }\bar{\psi}\gamma
^{\mu }\left( \delta \psi \right) \right] -m\bar{\psi}\left( \delta \psi
\right) \right\} \,,  \notag
\end{eqnarray}%
where $\mathcal{L}$ is the Lagrangian density in (\ref{lag-cv}).

Choosing different specializations of the variations, one can easily obtain,
for instance, the equations of motion and the energy-momentum and spin
tensors.

\subsection{Equations of Motion}

We choose to make functional variations only in the Dirac field thus we set $%
\delta \phi =\delta e^{\mu }{}_{b}=\delta \omega _{\mu ab}=0$ in the general
variation (\ref{action-variation}) and after some manipulations such as
integrations by parts, we obtain 
\begin{eqnarray}
\delta S &&\!\!=\!\!\!\int_{\Omega }dx~\partial _{\mu }\left[ \phi ^{4}e%
\frac{i}{2}\left( \bar{\psi}\gamma ^{\mu }\frac{1}{\phi }\left( \delta \psi
\right) -\frac{1}{\phi }\left( \delta \bar{\psi}\right) \gamma ^{\mu }\psi
\right) \right] ~\ +  \notag \\
&&  \label{eqsmov} \\[-0.2cm]
&&+\int_{\Omega }dx~\phi ^{4}e\left( \delta \bar{\psi}\right) \left[ i\gamma
^{\mu }\left( \nabla _{\mu }\psi +\frac{1}{2}\tilde{\tau}_{\mu }\psi \right)
-m\psi \right] +\int_{\Omega }dx~\phi ^{4}e\left[ -i\left( \nabla _{\mu }%
\bar{\psi}+\frac{1}{2}\tilde{\tau}_{\mu }\bar{\psi}\right) \gamma ^{\mu }-m%
\bar{\psi}\right] \delta \psi \,,\qquad  \notag
\end{eqnarray}%
where $\tilde{\tau}_{\mu }$ is the trace torsion and, it is given by 
\begin{equation}
\tilde{\tau}_{\mu }=\tilde{\tau}_{\mu \rho }{}^{\rho }=\frac{3}{\phi }%
\partial _{\mu }\ln \bar{\phi}~\ .  \label{traza-torsion}
\end{equation}

Now, by the action principle we get the generator of the variations of the
spinorial field 
\begin{equation}
G_{\delta \psi }=\int_{\partial \Omega }d\sigma _{\mu }\left[ \phi ^{4}e%
\frac{i}{2}\left( \bar{\psi}\gamma ^{\mu }\frac{1}{\phi }\left( \delta \psi
\right) -\frac{1}{\phi }\left( \delta \bar{\psi}\right) \gamma ^{\mu }\psi
\right) \right] \,,
\end{equation}%
and its equations of motion,%
\begin{eqnarray}
i\gamma ^{\mu }\left( \nabla _{\mu }\psi +\frac{1}{2}\tilde{\tau}_{\mu }\psi
\right) -m\psi  &=&0\,,  \notag \\
&& \\[-0.2cm]
i\left( \nabla _{\mu }\bar{\psi}+\frac{1}{2}\tilde{\tau}_{\mu }\bar{\psi}%
\right) \gamma ^{\mu }+m\bar{\psi} &=&0\,.  \notag
\end{eqnarray}

\subsection{Energy-Momentum Tensor and Spin Density Tensor}

Now, we vary only the background manifold and we assume that $\delta \omega
_{\mu ab}$ and $\delta e^{\mu }{}_{a}$ are independent variations, the
general variation (\ref{action-variation}) reads 
\begin{equation}
\delta S\;=\,\int_{\Omega }dx~\,\phi ^{4}e\left\{ \left[ \frac{i}{2}\left( 
\frac{{}}{{}}\bar{\psi}\gamma ^{a}\nabla _{\mu }\psi -\nabla _{\mu }\bar{\psi%
}\gamma ^{a}\psi \right) -e_{\mu }{}^{a}\mathcal{L}\right] \delta e^{\mu
}{}_{a}+\frac{i}{2}\left( \frac{{}}{{}}\bar{\psi}\gamma ^{\mu }\Sigma
^{ab}\psi +\bar{\psi}\Sigma ^{ab}\gamma ^{\mu }\psi \right) \frac{1}{2}%
\delta \omega _{\mu ab}\right\} \,,  \label{VarAcBack}
\end{equation}%
where we have used $\delta e=-ee_{\mu }{}^{a}\delta e^{\mu }{}_{a}$. First,
holding only the variations in the tetrad field, $\delta \omega _{\mu ab}=0$%
, we find 
\begin{equation}
\delta S=\int_{\Omega }dx\phi ^{4}e\left[ \frac{i}{2}\left( \bar{\psi}\gamma
^{a}\nabla _{\mu }\psi -\nabla _{\mu }\bar{\psi}\gamma ^{a}\psi \right)
-e_{\mu }{}^{a}\mathcal{L}\right] \delta e^{\mu }{}_{a}\,,
\end{equation}%
the expression between brackets defines the energy-momentum density tensor, 
\begin{equation}
T_{\mu }{}^{a}\equiv \frac{1}{\phi ^{4}e}\frac{\delta S}{\delta e^{\mu
}{}_{a}}=\frac{i}{2}\left( \bar{\psi}\gamma ^{a}\nabla _{\mu }\psi -\nabla
_{\mu }\bar{\psi}\gamma ^{a}\psi \right) -e_{\mu }{}^{a}\mathcal{L}\,.
\label{energy-momentum}
\end{equation}%
which can be written in coordinates as $T_{\mu }{}^{\nu }\equiv e^{\nu
}{}_{a}T_{\mu }{}^{a}$.

\vskip0.5cm Now, making functional variations only in the components of the
spin connection, $\delta e^{\mu }{}_{a}=0$, we get from (\ref{VarAcBack}) 
\begin{equation}
\delta S=\int_{\Omega }dx\phi ^{4}e\frac{i}{2}\left[ \frac{{}}{{}}\bar{\psi}%
\gamma ^{\mu }\Sigma ^{ab}\psi +\bar{\psi}\Sigma ^{ab}\gamma ^{\mu }\psi %
\right] \frac{1}{2}\delta \omega _{\mu ab},
\end{equation}%
and, we define the spin density tensor as being 
\begin{equation}
S^{\mu ab}\equiv \frac{2}{\phi ^{4}e}\frac{\delta S}{\delta \omega _{\mu ab}}%
=\frac{i}{2}\bar{\psi}\left\{ \frac{{}}{{}}\gamma ^{\mu },\Sigma
^{ab}\right\} \psi =\frac{1}{2}\varepsilon^{abc}{}_{d}\,e^{\mu }{}_{c}\bar{%
\psi}\gamma ^{5}\gamma ^{d}\psi ~,  \label{spin tensor}
\end{equation}
where we have used the relation $\ i\left\{ \gamma ^{c},\Sigma ^{ab}\right\}
=\varepsilon ^{abc}{}_{d}\gamma ^{5}\gamma ^{d}$.

\subsection{Functional Scale Variations and the Trace Relation}

Under pure infinitesimal variation of the scale function we get from the
equation (\ref{action-variation}) 
\begin{equation*}
\delta S=-\int_{\Omega }dx\phi ^{4}e\left[ \left( \frac{i}{2}\bar{\psi}%
\gamma ^{\mu }\nabla _{\mu }\psi -\frac{i}{2}\nabla _{\mu }\bar{\psi}\gamma
^{\mu }\psi -4\mathcal{L}\right) -\frac{i}{2}\bar{\psi}\left\{ \gamma ^{\mu
},S_{\mu }\right\} \psi \right] \frac{\delta \phi }{\phi }\,,
\end{equation*}%
and using the definition of the energy-momentum (\ref{energy-momentum}) and
the spin tensor (\ref{spin tensor}) we can write%
\begin{equation}
\delta S=-\int_{\Omega }dx~\phi ^{4}e\left( T_{\mu }{}^{a}e^{\mu }{}_{a}-%
\frac{1}{2}S^{\mu ab}\omega _{\mu ab}\right) \frac{\delta \phi }{\phi }\,,
\end{equation}%
from which we get the following identity:%
\begin{equation}
T_{\mu }{}^{a}e^{\mu }{}_{a}-\frac{1}{2}S^{\mu ab}\omega _{\mu ab}=0\,,
\label{trace-identity}
\end{equation}%
which is the so called trace relation. Such identity can be used to
constraint the form of the connection $\omega $ in a given content of matter.

\subsection{Conservation Laws}

As a final application of the Schwinger Action Principle, let us derive the
conservation laws associated to local Lorentz transformations and
infinitesimal general coordinate transformations.

\subsubsection{Spin Conservation}

Under \emph{local Lorentz transformations}, the functional variations of the
tetrad and the spin connection are given by%
\begin{equation}
\delta e^{\mu }{}_{a}=\delta \varepsilon _{a}{}^{b}e^{\mu }{}_{b}\,
\label{var-tetra}
\end{equation}%
\begin{equation}
\delta \omega _{\mu ab}=\omega _{\mu }{}^{c}{}_{b}\,\delta \varepsilon
_{ac}-\omega _{\mu {a}}{}^{c}\,\delta \varepsilon _{cb}-\frac{1}{\phi }%
\partial _{\mu }\delta \varepsilon _{ab}\,,  \label{var-spin}
\end{equation}%
with $\delta \varepsilon _{ab}=-\delta \varepsilon _{ba}$, where the first
variation expresses the vectorial character of the tetrad in Minkowsky
spacetime, and the second one comes from (\ref{TransfSpin}) with 
\begin{equation}
U=1+\frac{1}{2}\delta \varepsilon _{ab}\Sigma ^{ab}.
\end{equation}

The general expression (\ref{VarAcBack}) can be written using the
definitions of energy-momentum and spin tensors as being%
\begin{equation}
\delta S=\int_{\Omega}dx\phi^{4}e\left( T_{\mu}{}^{a}\delta
e^{\mu}{}_{a}+S^{\mu ab}\frac{1}{2}\delta\omega_{\mu ab}\right) \,.
\label{tgc-1}
\end{equation}

Substituting variations (\ref{var-tetra}) \ and (\ref{var-spin}) and after
some integration by parts and algebraic simplifications, we get%
\begin{equation}
\delta S=-\int_{\partial \Omega }d\sigma _{\mu }\left( \phi ^{3}eS^{\mu ab}%
\frac{1}{2}\delta \varepsilon _{ab}\right) +\int_{\Omega }dx\phi ^{4}e\left(
\nabla _{\mu }S^{\mu ab}+\tilde{\tau}_{\mu }S^{\mu ab}-T^{ab}+T^{ba}\right) 
\frac{1}{2}\delta \varepsilon _{ab}\,,
\end{equation}

Thus, from the Action Principle, we learn%
\begin{equation}
G_{\delta \varepsilon }=-\int_{\partial \Omega }d\sigma _{\mu }\left[ \phi
^{3}eS^{\mu ab}\frac{1}{2}\delta \varepsilon _{ab}\right] \,,
\label{spin-gen}
\end{equation}%
\begin{equation}
\nabla _{\mu }S^{\mu ab}+\tilde{\tau}_{\mu }S^{\mu ab}=T^{ab}-T^{ba}\,.
\label{ConservSpin}
\end{equation}%
$G_{\delta \varepsilon }$ is the generator of infinitesimal changes in the
spinorial field under local Lorentz transformations, and (\ref{ConservSpin})
gives us the conservation law for the spinning content of the theory. The
most important aspect here is the coupling among the spin density tensor and
the propagating torsion.

\subsubsection{Energy-Momentum Consevation}

Now we will concentrate our attention on the more complicated case of a 
\emph{general coordinate transformation}. From Lyra's transformation rule
for vectors and the infinitesimal displacement $\bar{x}^{\mu }=x^{\mu
}+\delta x^{\mu }$, we have,%
\begin{equation}
\bar{e}^{\mu }{}_{a}\left( \bar{x}\right) =\frac{\bar{\phi}\left( \bar{x}%
\right) }{\phi \left( x\right) }\frac{\partial \bar{x}^{\mu }}{\partial
x^{\nu }}e^{\nu }{}_{a}\left( x\right) \,,
\end{equation}%
with its functional variation $\delta e^{\mu }{}_{a}\left( x\right) \equiv 
\bar{e}^{\mu }{}_{a}(x)-e^{\mu }{}_{a}\left( x\right) $ given by%
\begin{equation}
\delta e^{\mu }{}_{a}\left( x\right) =\frac{\bar{\phi}\left( x\right) }{\phi
\left( x\right) }\left[ e^{\mu }{}_{a}\left( x\right) +e^{\nu
}{}_{a\,}\left( x\right) \partial _{\nu }\delta x^{\mu }-\delta x^{\nu
}\partial _{\nu }e^{\mu }{}_{a}\left( x\right) +e^{\mu }{}_{a}\left(
x\right) \delta x^{\nu }\partial _{\nu }\ln \phi \left( x\right) \right]
-e^{\mu }{}_{a}\left( x\right) \,.  \label{gct-tetrad}
\end{equation}%
While for the covariant Lyra vector defined by the spin connection we have 
\begin{equation}
\bar{\omega}_{\mu ab}^{\,}\left( \bar{x}\right) =\frac{\phi \left( x\right) 
}{\bar{\phi}\left( \bar{x}\right) }\frac{\partial x^{\nu }}{\partial \bar{x}%
^{\mu }}\omega _{\nu ab}\left( x\right) \,,
\end{equation}%
and its functional variation is%
\begin{equation}
\delta \omega _{\mu ab}\left( x\right) =\frac{\phi \left( x\right) }{\bar{%
\phi}\left( x\right) }\left[ \omega _{\mu ab}\left( x\right) -\omega _{\nu
ab}\left( x\right) \partial _{\mu }\delta x^{\nu }-\delta x^{\nu }\partial
_{\nu }\omega _{\mu ab}\left( x\right) -\omega _{\mu ab}\left( x\right)
\delta x^{\nu }\partial _{\nu }\ln \phi \left( x\right) \right] -\omega
_{\mu ab}\left( x\right) \,.  \label{gct-spinc}
\end{equation}

Substituting the expressions (\ref{gct-tetrad}) and (\ref{gct-spinc}) in (%
\ref{tgc-1}) and, after some calculations we obtain for the variation of the
action the following expression%
\begin{eqnarray}
\delta S &=&\!\int_{\Omega }d\sigma _{\mu }~e\phi ^{4}\left[ \frac{\bar{\phi}%
}{\phi }T_{\nu }{}^{\mu }-\frac{\phi }{\bar{\phi}}S^{\mu ab}\frac{1}{2}%
\omega _{\nu ab}\right] \delta x^{\nu }+\int_{\Omega }dx~e\phi ^{4}\left( 
\frac{\bar{\phi}}{\phi }-1\right) \left( T_{\mu }{}^{a}e^{\mu }{}_{a}-\frac{%
\phi }{\bar{\phi}}\frac{1}{2}\omega _{\mu ab}S^{\mu ab}\right)  \notag \\
&&-\int_{\Omega }dx~e\phi ^{5}\left[ \frac{\bar{\phi}}{\phi }\left( \nabla
_{\mu }T_{\nu }{}^{\mu }+\tilde{\tau}_{\mu \nu }{}^{\lambda }T_{\lambda
}{}^{\mu }+\tilde{\tau}_{\mu }T_{\nu }{}^{\mu }\right) +\frac{\bar{\phi}}{%
\phi ^{2}}\left( \partial _{\mu }\ln \frac{\bar{\phi}}{\phi }\right) T_{\nu
}{}^{\mu }+\right.  \label{coordenada} \\
&&\left. ~\ \ \ \ \ \ \ \ \ \ \ \ \ \ \ \ +\left( \frac{\bar{\phi}}{\phi }-%
\frac{\phi }{\bar{\phi}}\right) \omega _{\nu ab}T^{ab}+\frac{1}{\bar{\phi}}%
\left( \partial _{\mu }\ln \frac{\bar{\phi}}{\phi }\right) \frac{1}{2}\omega
_{\nu ab}S^{\mu ab}+\frac{1}{2}\frac{\phi }{\bar{\phi}}S^{\mu ab}R_{\nu \mu
ab}\right] \delta x^{\nu }.  \notag
\end{eqnarray}

According to the Schwinger Principle, from the surface integral we obtain
the generator of the infinitesimal changes in the Dirac field under general
coordinate transformations 
\begin{equation}
G_{\delta x}=\int_{\Omega }d\sigma _{\mu }~e\phi ^{4}\left[ \frac{\bar{\phi}%
}{\phi }T_{\nu }{}^{\mu }-\frac{\phi }{\bar{\phi}}\frac{1}{2}\omega _{\nu
ab}S^{\mu ab}\right] \delta x^{\nu }\,.  \label{1ra-1}
\end{equation}

Next, from the third integral in (\ref{coordenada}) and due to the
invariance of the action under general coordinate transformations we obtain
the conservation law of energy-momentum. 
\begin{gather}
\frac{\bar{\phi}}{\phi }\left( \nabla _{\mu }T_{\nu }{}^{\mu }+\tilde{\tau}%
_{\mu \nu }{}^{\lambda }T_{\lambda }{}^{\mu }+\tilde{\tau}_{\mu }T_{\nu
}{}^{\mu }\right) +\frac{1}{2}\frac{\phi }{\bar{\phi}}S^{\mu ab}R_{\nu \mu
ab}+  \notag \\
+\left( \frac{\bar{\phi}}{\phi }-\frac{\phi }{\bar{\phi}}\right) \omega
_{\nu ab}T^{ab}+\frac{1}{\bar{\phi}}\left( \partial _{\mu }\ln \frac{\bar{%
\phi}}{\phi }\right) \left[ \frac{\bar{\phi}^{2}}{\phi ^{2}}T_{\nu }{}^{\mu
}+\frac{1}{2}S^{\mu ab}\omega _{\nu ab}\right] \;=\;0\,.
\end{gather}

By imposing the invariance of the action under general coordinate
transformations, the second integral in (\ref{coordenada}) allows to get a
new relation, we named it as the trace symmetry. Such identity for $\bar{\phi%
}\neq \phi $ can be written as 
\begin{equation}
T_{\mu }{}^{a}e^{\mu }{}_{a}-\frac{\phi }{\bar{\phi}}\frac{1}{2}S^{\mu
ab}\omega _{\mu ab}=0\,,  \label{2da-3}
\end{equation}%
it can be considered as the generalization of the trace relation shown in (%
\ref{trace-identity}). Formally, for $\bar{\phi}\equiv \phi $ it reduces to (%
\ref{trace-identity}).

If we select the special case $\bar{\phi}\equiv \phi $, we get the following
equations 
\begin{equation}
G_{\delta x}=\int_{\Omega }d\sigma _{\mu }e\phi ^{4}\left( T_{\nu }{}^{\mu
}-S^{\mu ab}\frac{1}{2}\omega _{\nu ab}\right) \delta x^{\nu }\,,
\end{equation}

\begin{equation}
\nabla _{\mu }T_{\nu }{}^{\mu }+\tilde{\tau}_{\mu \nu }{}^{\lambda
}T_{\lambda }{}^{\mu }+\tilde{\tau}_{\mu }T_{\nu }{}^{\mu }+\frac{1}{2}%
S^{\mu ab}R_{\nu \mu ab}=0\,,  \label{EnerMomFi1}
\end{equation}

\section{Final Remarks}

We have constructed the coupling of spinorial fields to Lyra background and
found the general laws governing the motion of the spinor and the
conservation of energy-momentum and spin, using several specializations of
the Schwinger Action Principle which were applied in a classical context.

It is possible to establish the Riemannian limit of the Lyra geometry
setting the scale functions to be $\phi =\bar{\phi}\equiv 1$, thus, the
conservation laws for the Dirac fields reads%
\begin{equation}
\nabla _{\mu }S^{\mu ab}=T^{ab}-T^{ba}\,.
\end{equation}%
and%
\begin{equation}
\nabla _{\mu }T_{\nu }{}^{\mu }+\frac{1}{2}S^{\mu ab}R_{\nu \mu ab}=0\,,
\end{equation}

The conservation laws (\ref{EnerMomFi1}) and (\ref{ConservSpin}) for the
Dirac fields, in the case $\phi =\bar{\phi}$, are similar to the laws
founded when the background is the Riemann-Cartan geometry. Such
identification can be done by replacing $\tilde{\tau}_{\mu \nu }{}^{\lambda
} $ $\rightarrow 2Q^{\mu }{}_{\beta \nu }$, where $Q^{\mu }{}_{\beta \nu
}=\Gamma ^{\mu }{}_{\left[ \beta \nu \right] }$ is the RC torsion, thus we
have spin conservation law 
\begin{equation}
\nabla _{\mu }S^{\mu ab}+2Q_{\mu }S^{\mu ab}=T^{ab}-T^{ba}
\end{equation}%
and the energy-momentum law 
\begin{equation}
\nabla _{\mu }T_{\nu }{}^{\mu }+2Q^{\mu }{}_{\beta \nu }\,T_{\mu }{}^{\beta
}+2Q_{\mu }T_{\nu }{}^{\mu }+\frac{1}{2}S^{\mu ab}R_{\nu \mu ab}=0
\end{equation}%
where $Q_{\mu }=Q^{\nu }{}_{\mu \nu }$. Such as obtained in \cite{Nieh-1}.

On the other hand, the trace symmetry (\ref{2da-3}) and the trace relation (%
\ref{trace-identity}) are new properties provided by the scale invariance of
the Lyra geometry. The results here obtained are still preliminaries, but it
seems to indicate that Lyra Geometry would be a useful tool to implement a
kind of conformal symmetry even in the case of massive fields. To make
precise this reason, it is necessary a deep study of the relation among
conformal invariance and Lyra scale transformations. The investigation of
the back reaction of the one-half spin field upon the geometry would show us
how a propagation equation for the scale function can be obtained. These
studies are currently under discussion.

\begin{center}
\textbf{Acknowledgments}
\end{center}

This work is supported by FAPESP grants 01/12611-7 (RC), 01/12584-0 (CAMM)
and 02/00222-9 (BMP). BMP also thanks CNPq for partial support.

\end{document}